# Multi-mJ, kHz, 2.1-μm OPCPA for high-flux soft X-ray high-harmonic radiation

Kyung-Han Hong,[1,*] Chien-Jen Lai,[1] Jonathas Siqueira,[1,2] Peter Krogen,[1] Jeffrey Moses,[1] Martin Smrz,[1] Luis E. Zapata,[1] and Franz X. Kärtner[1,3]

[1]*Department of Electrical Engineering and Computer Science and Research Laboratory of Electronics, Massachusetts Institute of Technology (MIT), Cambridge, Massachusetts 02139, USA*
[2] *Instituto de Fisca de Sao Carlos, SP 13560-970, Brazil*
[3]*Center for Free-Electron Laser Science, DESY and Department of Physics, University of Hamburg, Hamburg, Germany*
*\*Email: kyunghan@mit.edu*

**Abstract:** We report on a multi-mJ 2.1-μm OPCPA system operating at a 1-kHz repetition rate, pumped by a picosecond cryogenic Yb:YAG pump laser, and the phase-matched high-flux high-harmonic soft X-ray generation.

## 1. Introduction

High-order harmonic generation (HHG) with long-wavelength drive pulses [1] has been proven to be the most reliable way of extending the photon energy of coherent attosecond harmonic pulses to water-window soft X-ray (280-520 eV) [2,3] and even hard X-ray regions [4]. However, since the HHG efficiency in the single-atom response unfavorably scales with driving wavelength, i.e., $\lambda^{-(5-6)}$, mostly governed by quantum diffusion of electron trajectories, the phase matching between the driving pulse and the generated harmonics is very important for avoiding an additional efficiency drop. In general, multi-mJ pulse energy is required to avoid the phase mismatch induced by the Guoy phase shift. High gas pressures in the interaction region are also important for the phase matching in Ne and He atoms used for the largest cutoff extension. Optical parametric amplification (OPA) and optical parametric chirped pulse amplification (OPCPA) techniques have enabled to generate long-wavelength driving pulses at 1.3-4 μm. Recently, phase-matched HHG with an efficiency of $\sim 10^{-8}$ covering the water window range was experimentally demonstrated [3] using a 10-Hz, 2.4-mJ, 2-μm OPA source pumped by a Ti:sapphire laser. However, the generated XUV photon number per second is still as low as $\sim 6 \times 10^7$ over the entire water window range, limiting potential applications. Increasing the repetition rate to kHz is a clear approach for increasing soft X-ray photon flux by orders of magnitude. Ishii *et al.* [5] used a kHz multi-mJ, few-cycle 750-nm OPCPA system to generate soft X-ray harmonics with a cutoff of ~300 eV just into the water window region, but the efficiency sharply dropped even before the cutoff while the photon flux was not reported.

In this paper, we report on a kHz multi-mJ OPCPA at 2.1 μm for the phase-matched HHG in the water window region towards a flux of $\sim 10^8$ photons/s/1% bandwidth (or $\sim 10^{10}$ over the entire water-window region). First, we describe the development of an OPCPA pump laser based on a kHz, picosecond cryogenic Yb:YAG amplifier generating up to 42 mJ after compression. Second, we scale up the energy of our OPCPA system from 0.85 mJ [6] to 2.6 mJ using this pump laser. And then, we will discuss the results of soft X-ray HHG driven by this high-energy 2.1-μm source.

## 2. OPCPA pump laser based on cryogenic Yb:YAG multi-stage chirped-pulse amplifiers

Although the OPCPA technology is generally accepted as a more energy-scalable method than OPA, energy scaling at high repetition rates in the kHz range is a nontrivial task without a suitable picosecond pump source. Since such task was ultimately limited by the picosecond Nd:YLF laser technology, new pump laser technologies based on Yb:YAG gain medium have been explored. Metgzer *et al.* [7] developed a 25-mJ ~1-ps thin-disk Yb:YAG amplifier at 3 kHz, which recently enabled the generation of 1.2 mJ, 2.1 μm pulses at 3 kHz [8]. In the other hand, Hong *et al.* [9] have demonstrated 40 mJ of amplification of uncompressed pulses at 2 kHz from a picosecond cryogenic Yb:YAG chirped pulse amplifier (CPA). A modified cryogenic Yb:YAG laser with 15 mJ of compressed output energy was used for pumping a 0.85-mJ, 2.1-μm OPCPA system [6]. In this paper, we further upgrade this pump laser using an additional amplifier stage and generate 42-mJ, 17-ps pulses (56 mJ of energy before compression) at a 1-kHz repetition rate.

The system configuration is illustrated in Fig. 1(a). The seed from the Ti:sapphire oscillator, which also provides the seed for the OPCPA stages, is stretched by a chirped volume Bragg grating (CVBG) pair to ~560 ps with 0.7 nm of bandwidth at 1029 nm, and then amplified by a regenerative amplifier and a two-stage multipass amplifier that are both based on cryogenically cooled Yb:YAG as gain medium. We used a 1% doped 10-mm-long Yb:YAG crystal in the first (2-pass) amplifier and a 2% doped 20-mm-long crystal in the second (single-pass) amplifier. Each crystal is cooled by liquid nitrogen in a vacuum chamber connected to liquid nitrogen auto-refilling systems. The maximum energies from the first and second amplifiers are 30 and 56 mJ, respectively, with excellent beam profiles as shown in Fig. 1(b) and (c) where the fringes are measurement artifacts. The pump powers from two 940-nm laser diodes at the maximum output energy are both ~240 W. The amplified pulse is compressed to 17 ps using a multi-layer dielectric grating pair with a throughput efficiency of 75%, delivering a maximum compressed energy of 42 mJ. To our best knowledge, this is the highest energy demonstrated from a picosecond laser system at kHz repetition rates.

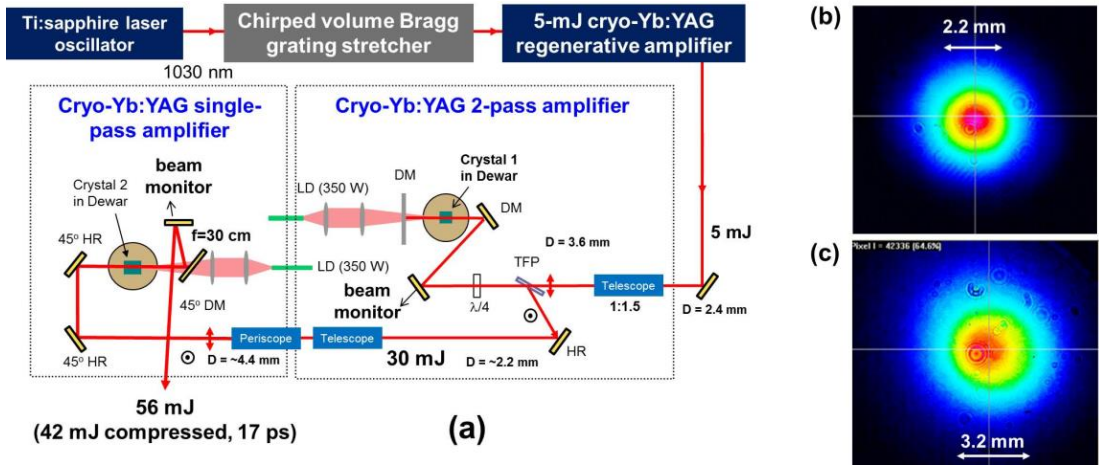

**Fig. 1** High-energy picosecond cryogenic Yb:YAG pump laser operating at 1 kHz repetition rate composed of a regenerative amplifier and 2-stage multipass amplifier (a). The near-field beam profiles from the first 2-pass amplifier (b) and the second single-pass amplifier (c) show a near-Gaussian shape, where the fringes are measurement artifacts.

**3. 2.1-μm OPCPA and soft X-ray HHG**

In the previous reports [6,10], we have developed a kHz, 0.85-mJ, 2.1-μm ultrabroadband OPCPA system, pumped by two independent picosecond lasers: 1) 1047 nm Nd:YLF amplifier and 2) 1029 nm cryogenic Yb:YAG amplifier. For the multi-mJ 2.1-μm pulse amplification, we maintain the same configuration except for the much higher energy from the Yb:YAG pump laser, the main power pump source. The schematic diagram of the OPCPA system is depicted as Fig. 2. Both pump lasers are seeded by one Ti:sapphire laser for achieving optical synchronization. The Nd:YLF chirped-pulse amplifier (CPA) pumps the first two pre-OPCPA stages. The required and used pump energy for the first two stages of the OPCPA was 1.9 mJ out of 4 mJ. The 42-mJ cryogenic Yb:YAG laser pumps the final OPCPA stage for power amplification.

    The ultrabroadband 2.1-μm pulse is generated from the intrapulse difference-frequency generation (DFG) and stretched by an anti-reflection (AR) coated silicon (Si) block (~5 ps) and AOPDF (~9 ps) before being amplified in the first and second OPA stages. Ultrabroadband amplification is achieved by means of the degenerate OPA in PPLN and PPSLT crystals. The details of the first two OPA stages are described in Ref. [10]. The ~25 μJ pulse from the second OPA stage is further stretched to ~17 ps by another AR coated Si block and then amplified in a 5-mm-long type I BBO crystal. After optimization in the third OPA stage, we obtained a maximum energy of 2.6 mJ with good spatial beam quality at the third stage with a gain of ~100 at a pump intensity of 30-50 GW/cm$^2$. To minimize the pump beam instability due to air fluctuations, we installed a pointing stabilizer for the Yb:YAG pump beam. The spectral bandwidth is 370 nm centered at 2.1 μm and the compressed pulse duration is ~45 fs, measured using an interferometric autocorrelator. It should be noted that the carrier-envelope phase (CEP) of the 2.1-μm pulses is passively stabilized [10], but the CEP stability was not measured because of the long pulse duration of 6.5 cycles.

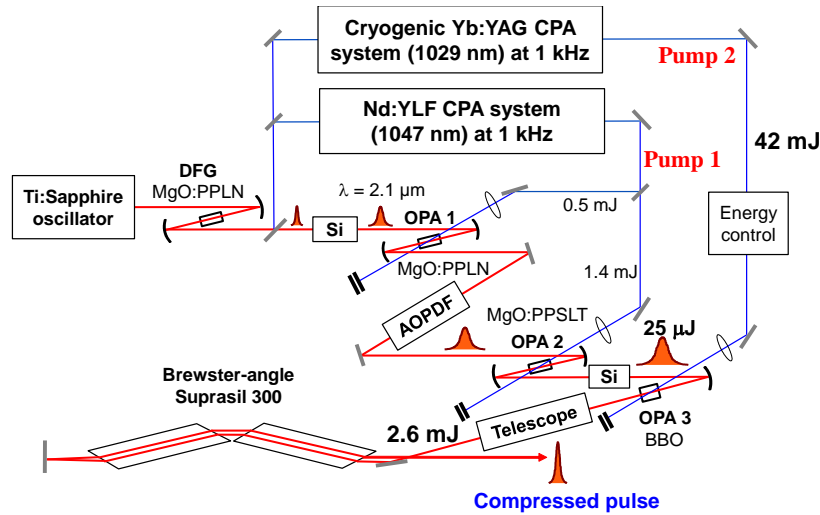

**Fig. 2** Optical layout of the ultrabroadband 2.1-μm 3-stage OPCPA system.

We delivered the 2.1-μm pulses into a HHG chamber and focused the pulses at a gas target. Figure 3 shows an example of 2.1-μm-driven HHG [11] in an Ar gas jet with a pressure of 70 mbar at the interaction region. We demonstrated a cutoff extension to 160 eV in Ar with a photon flux of $0.8 \times 10^8$ photons/s/over 1% bandwidth at 130 eV using 0.4-mJ driving energy. Since the gas jet does not support pressures of Ne and He high enough for phase matching in the water window region, we have developed a gas cell as well as a hollow fiber target. These target systems were tested for HHG with a Ti:sapphire laser amplifier. Currently, 2.1-μm-driven HHG experiments with Ne and He gases using both the gas cell and the hollow fiber are in progress. The cutoff extension to the water window region with a flux of more than $10^8$ photons/s/over 1% bandwidth is expected. The detailed characterization of the high-flux soft X-ray HHG will be presented.

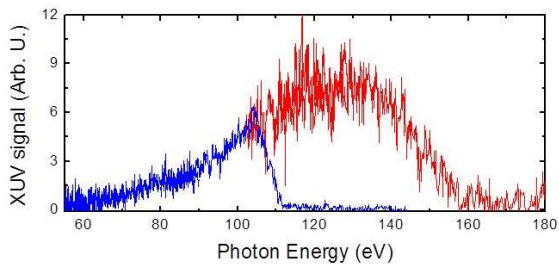

**Fig. 3** Experimentally measured HHG spectrum from Ar gas jet. The blue and red curves show the measurements with Be and Zr filters, respectively. (pressure of ~70 mbar, intensity of $(1.7 \pm 0.2) \times 10^{14}$ W/cm$^2$, efficiency of ~$2 \times 10^{-9}$ at ~130 eV, XUV flux: $0.8 \times 10^8$ photons/s/1% bandwidth)

In summary, we developed a multi-mJ, kHz, 2.1-μm OPCPA, pumped by a 42-mJ picosecond cryogenic Yb:YAG amplifier, which is suitable for the high-flux phase-matched soft X-ray HHG in the water-window region.

## 4. References
[1] A. Gordon and F. X. Kärtner, Opt. Express **13**, 2941 (2005).
[2] E. J. Takahashi, T. Kanai, K. L. Ishikawa, Y. Nabakawa, and K. Midorikawa, Phys. Rev. Lett. **101**, 253901 (2008).
[3] M. C. Chen *et al.*, Phys. Rev. Lett. **105**, 173901 (2010)
[4] T. Popmintchev *et al.*, Science **336**, 1287 (2012).
[5] N. Ishii, S. Adachi, Y. Nomura, A. Kosuge, Y. Kobayashi, T. Kanai, J. Itatani, and S. Watanabe, Opt. Lett **37**, 97 (2012).
[6] K.-H. Hong, S.-W. Huang, J. Moses, X. Fu, C.-J. Lai, G. Cirmi, A. Sell, E. Granados, P. Keathley, and F. X. Kärtner, Opt. Express **19**, 15538-15548 (2011).
[7] T. Metzger, A. Schwarz, C. Y. Teisset, D. Sutter, A. Killi, R. Kienberger, and F. Krausz, Opt. Lett. **34**, 2123 (2009).
[8] Y. Deng *et al.*, Opt. Lett. **37**, 4973 (2012).
[9] K.-H. Hong, J. Gopinath, D. Rand, A. Siddiqui, S.-W. Huang, E. Li, B. Eggleton, John D. Hybl, T. Y. Fan and F. X. Kärtner, Opt. Lett. **35**, 1752 (2010).
[10] J. Moses, S.-W. Huang, K.-H. Hong, O. D. Mücke, E. L. Falcão-Filho, A. Benedick, F. Ö. Ilday, A. Dergachev, J. A. Bolger, B. J. Eggleton and F. X. Kärtner, Opt. Lett. **34**, 1639 (2009).
[11] K.-H. Hong, C.-J. Lai, V.-M. Gkortsas, S.-W. Huang, J. Moses, E. Granados, S. Bhardwaj, and F. X. Kärtner, Phys. Rev. A **86**, 043412 (2012).